\documentclass{ws-procs9x6}

\begin{document}

\title{Summary of the Heavy Flavours Working Group}

\author{U. Karshon$^*$, I. Schienbein$^{**}$ and P. Thompson$^{***}$}

\address{$^*$Weizmann Institute of Science, Israel\\
$^{**}$Southern Methodist University, Dallas, USA\\
$^{***}$University of Birmingham, UK}  

%\author{U. Karshon}
%\address{Weizmann Institute of Science, Israel}

%\author{I. Schienbein}
%\address{Southern Methodist University, Dallas, USA}  

%\author{P. Thompson}
%\address{University of Birmingham, UK}  

\maketitle

\abstracts{
This is a summary of the contributions presented in the Heavy Flavours Working
Group of the DIS2006 Workshop.
}

\section{Introduction and Overview}
\label{sec:intro}

Heavy flavour physics is an ongoing active field of experimental and theoretical research.
As in all previous DIS Workshops, the emphasize in this Working Group was on the
production of the heavy quarks charm (c) and beauty (b) and on issues related to
QCD. We held five pure Heavy Flavours sessions, two joint sessions with the Hadronic
Final States Working Group  and one joint session with the Structure Functions and
Low-$x$ Working Group.  
In total, there were 23 experimental talks and 12
theoretical ones in all the above sessions.
%There were about twice as many experimental talks (23) compared to the
%theoretical ones (12) in all the above sessions.
 
The many presentations from the two collider HERA experiments H1 and ZEUS were
complemented by significant contributions from the Tevatron at Fermilab, the
B-factories Babar and Belle and the RHIC heavy ion collider at Brookhaven.
The main topics discussed in the Heavy Flavours Working Group were: heavy flavour
schemes and fragmentation, heavy quark production at RHIC, open charm and beauty
production at HERA, quarkonium production, beauty production at the Tevatron, charm
spectroscopy, charm decays and new states at the B-factories. The joint session with
the Structure Functions Working Group concentrated in heavy quark structure functions
and parton density functions. The joint sessions with the Hadronic Final States Working
Group dealt with common theory issues, such as $k_T$ factorization schemes and jet
definitions and with searches for exotic baryonic states (pentaquarks).
 
Following the presentation of the summary talks at the Workshop, 
the remainder of this document is divided into two
parts, a theory and an experimental summary.
%, concluded by some final comments.

\section{Theory summary}
\label{sec:theory}
In this section, we summarize the main theoretical issues that
were presented in the Heavy Flavours Working Group of DIS 2006.
Several theoretical talks have been related 
to the treatment of heavy quarks in perturbation theory
('heavy flavour schemes'). 
In particular, we had discussions on heavy flavour schemes for the fully
inclusive case used in global analyses of parton distribution functions 
($\alpha_s$ treatment, NNLO, S-ACOT($\chi$), intrinsic charm).
Moreover, we had talks related to heavy flavour schemes in one-particle
inclusive processes relevant for global analyses of fragmentation functions
and a proper description of kinematic distributions in the momentum of the
observed heavy-flavoured hadron (GM-VFNS, evolution with heavy quark thresholds).
Other major topics have been 
a new analysis of heavy quark fragmentation functions, 
heavy quark production at RHIC, progress on heavy quark production
in the $k_T$-factorization approach and a comparison of charmonium
production in the colour evaporation model (CEM) and non-relativistic
QCD (NRQCD).

\subsection{Heavy Flavour Schemes}
Heavy flavour schemes have been studied intensively in the
past three decades.
At the heart of these investigations 
is the question of how to deal with heavy quarks in perturbative QCD.
%
% Define FFN,ZM-VFN,GM-VFN
In the fixed flavour number scheme (FFNS), the heavy quark is treated
in fixed order (FO) perturbation theory, i.e., collinear logarithms 
of the heavy quark mass are computed order by order in perturbation theory.
On the other hand, in a variable flavour number scheme (VFNS)
these collinear logarithms are absorbed
into heavy quark parton distribution functions (PDFs) and fragmentation
functions (FFs) at (or close to) the heavy quark mass scale $m_h$.
The logarithms are resummed to all orders in perturbation theory
by evolving from $m_h$ to the hard scale $Q$ of the process
with the help of the well-known DGLAP renormalization group equations.
The number of active flavours, $n_f$, is 'variable' because
at the scale $m_h$, where the heavy quark PDF is introduced, $n_f$ 
is increased by one unit.
There are several realizations of variable flavour number schemes
discussed in the literature due to the fact that the treatment
of finite $m_h/Q$ terms is not prescribed by the factorization and 
renormalization schemes.
In the Zero-Mass VFNS (ZM-VFNS) the finite heavy quark mass terms are neglected. 
On the other hand, General-Mass VFNS (GM-VFNS) have prescriptions
how to take into account the $m_h/Q$ pieces which is essential for
applications which include regions of phase space where $Q \sim m_h$,
such as global fits to determine PDFs.

% Of general interest for the user

% Still ongoing work.
% Sometimes technical but important to understand certain
% features of the development for the 'user' of PDFs/FFs
% in order to perform consistent calculations.

\subsubsection{Fully Inclusive Case\protect\cite{thorne,tung}}
R.\ Thorne, in his contribution, has emphasized the 
importance of using a GM-VFNS for performing a 
fully-global analysis of PDFs.
Here, the term 'fully-global analysis' is opposed to 
semi-global fits which only include a part of the available 
experimental information.
%
% because of quality of fit but also since many processes
% are not calculated in the FFNS in NLO. More: FFN not
% worked out at NNLO
% m/Q terms not to be neglected (many data) -> no Zm_VFN
Nevertheless, there are several applications which require the
knowledge of PDFs in a FFNS. The MRST group now provides
such parton distributions\cite{Martin:2006qz} by evolving from the MRST04 partons
at $Q_0 = 1$ GeV but keeping $n_f=3$ fixed in the splitting functions
{\em and} $\alpha_s$. 

Another subject of his talk has been the construction of
a GM-VFNS at NNLO which is used in a global analysis of PDFs at NNLO\cite{Thorne:2006qt}.
So far, this is the only existing detailed proposal for a GM-VFNS at NNLO
and it should be stressed again that the
development of such a scheme is mandatory for global parton analyses 
at this order.

W.-K.\ Tung has reported on a new implementation of 
a heavy flavour scheme based on the
GM-VFNS of Collins\cite{Collins:1998rz} in the CTEQ
{\tt Fortran} package.
This scheme, the  S-ACOT($\chi$) scheme, combines the kinematic 
constraints of the ACOT($\chi$) rescaling procedure\cite{Tung:2001mv} with 
the  simplifications of the S-ACOT scheme\cite{Kramer:2000hn} 
which states that it is a convenient {\em scheme choice} to 
neglect the heavy quark mass terms in
all processes with a heavy quark in the initial state.
It should be noted that heavy quark initiated contributions
including the full heavy quark mass dependence have been computed
in next-to-leading order (NLO) for deep inelastic scattering
in Ref.\ \cite{Kretzer:1998ju}. However, similar calculations for other 
processes have not been performed.
This new implementation has been utilized in a new global analysis
of parton distributions, incorporating the complete HERA I data sets 
as described in the Structure Functions session summary.

Conventional global fits assume that heavy quark PDFs are 
{\em radiatively generated} by QCD evolution, fully relying on
perturbatively computed boundary conditions\cite{Buza:1998wv}.
Although
the assumption of a purely perturbative charm or bottom PDF does not contradict
any experimental data it is important to test it in a quantitative way.
While the possible amount of intrinsic charm has been analysed before
in the literature, see e.g. Ref.\ \cite{Harris:1996jx}, 
it has never been done in the context of a global
analysis which is the appropriate approach since the various quark
and gluon PDFs are intimately linked.
Within the framework of the new implemented S-ACOT($\chi$) scheme,
the CTEQ collaboration has included a non-perturbative charm distribution
at the input scale $Q=m_c$ and has performed a global analysis\cite{tung}.
Their results show that
the best fit favours a small non-perturbative charm component
at the input scale, although the difference between this and a conventional
fit with no intrinsic charm falls well within the uncertainty range of the
global fit. Furthermore, the magnitude of the intrinsic charm component
is limited to carry at most $1.8\%$ of the proton's momentum at $Q=m_c$. 

%Discussion: 
%NNLO (technically complicated, conceptionally clear) 
%or better understand resummation+correct implementation of schemes
%in NLO?

\subsubsection{One-Particle Inclusive Case\protect\cite{kniehl,cacciari1}}
For the same reasons as in the fully inclusive case it is
important to work out the details of a GM-VFNS for one-particle inclusive
production of heavy quarks/hadrons.
B.\ Kniehl reported on progress in the development of such
a scheme\cite{Kniehl:2004fy,Kniehl:2005mk}.
%
% Description
% As in the inclusive case collinear logarithms of the heavy quark mass
% are absorbed into heavy quark PDFs.
% 
Employing universal fragmentation functions for 
$D^0$, $D^+$, $D_s^+$, and
$D^{\star+}$ mesons (and complex conjugates)\cite{ffs}
extracted from a fit to LEP1 data from the OPAL
collaboration, a good description of CDF run II data could
be achieved\cite{Kniehl:2005ej}.
More results in this GM-VFNS are expected in the future for $B$-meson
production at the Tevatron and heavy-flavoured hadron production
in deep inelastic scattering at HERA.

%\subsubsection{Evolution of FFs with heavy quark thresholds 
%\protect\cite{cacciari1}}
Another topic was the evolution of FFs across heavy quark thresholds\cite{cacciari1}.
%Similar to parton distribution functions, fragmentation functions 
%change when the charm or bottom threshold is crossed.
While the proper perturbative relations ('matching conditions') between 
the parton distributions below and above the heavy quark thresholds are known 
and used in NLO QCD since 20 years\cite{Collins:1986mp}
this problem has been ignored so far in fits to light
hadron fragmentation functions.
In his talk, M.\ Cacciari discussed the relevant NLO matching conditions
for fragmentation functions which should be used in the evolution 
across heavy flavour thresholds\cite{Cacciari:2005ry} in
cases where the heavy flavour FFs are radiatively generated.
%(by QCD evolution of perturbative initial conditions).
%
%Analogously to the parton distribution function case,
%the heavy quark fragmentation functions are 
%generated from the perturbative initial conditions
%when evolving through the thresholds.
%%
%Note however, that the initial conditions for the heavy quark FFs
%differ from the corresponding initial conditions for the charm and bottom PDFs
%which vanish at the mass thresholds $\mu = m_c$ resp. $\mu = m_b$.
%
As an interesting application, 
it will be
possible to perform fits to light hadron fragmentation data parametrizing
only the three light quarks and the gluon FFs while
the charm and bottom FFs are purely perturbative.
%Since the charm and bottom FFs are radiatively generated 
%(i.e., from {\em perturbative initial conditions} 
%using the Altarelli-Parisi evolution equations)
%it will be
%possible to perform fits to light hadron fragmentation data parametrizing
%only the three light quarks and the gluon FFs.

\subsection{Heavy Quark Fragmentation\protect\cite{oleari}}
C.\ Oleari gave a theoretical introduction to heavy quark fragmentation
functions.
%(for details of the introduction see the transparencies).
Moreover, he reported on results of a recent QCD analysis 
of $D$ and $B$ meson FFs\cite{Cacciari:2005uk}
using Belle and CLEO data at $\sqrt{S} = 10.6$ GeV and LEP1 data
at $\sqrt{S}=m_Z$.
%
% Important study: 
This study is interesting for at least two reasons.
First, the $D$-meson data from the $B$-factories are very precise
and, secondly, the two different energy scales allow to test the
universality of the fragmentation functions. 
% in analogy of universality of PDFs/scaling violations of F_2(x,Q^2)
%
The analysis was performed in NLO QCD (NLO initial conditions, evolution, and
coefficient functions) including 
soft gluon resummation effects at the next-to-leading-log (NLL) level,
evolution with proper matching at the bottom threshold (as discussed in the previous section)
and correcting the data for QED initial state radiation. 
%
% There are two remarkable results:
% I.
As a result, the description of CLEO and Belle data for $D$ mesons was very good in the
whole $x$-range.
% Noteworthy, in contrast to earlier studies
% II.
However, evolving the fit of CLEO and Belle data to the $Z$-pole resulted in a
bad description of ALEPH data in the large-$x$ (large-$N$) region
questioning the universality of the fragmentation functions.
Note that the analogue in the space-like sector would be
PDFs determined from data of the
structure function $F_2(x,Q^2)$ at low-$Q^2$ which,
after DGLAP evolution to large $Q^2$, result in a bad description of 
the structure function data at the high scale. 
In this case one would probably think of target mass corrections
or higher twist effects as possible explanations of the discrepancy.
For the fragmentation functions, a possible explanation\cite{Cacciari:2005uk} of this 
worrisome result could be non-perturbative corrections to the
coefficient function of type $1+C (N-1)/Q^2$ or $1+C (N-1)/Q$. 
Clearly, more work is needed here in the future.

\subsection{Heavy Quark Production at RHIC\protect\cite{cacciari2}}
M.\ Cacciari has presented QCD benchmark predictions for
open charm and bottom production at RHIC\cite{Cacciari:2005rk}
based on the FONLL framework\cite{Cacciari:1998it} supplemented with suitable
spectra for the decay of the heavy flavoured hadron $H_Q$ into the
observed electron.
The corresponding theoretical uncertainties are discussed in detail
which should be an important part of any reliable theoretical study.
As a result, within errors the theoretical predictions are in
fair agreement with the RHIC data
for $pp$ and $d Au$ collisions at $\sqrt{S_{NN}}=200$ GeV. 
However, the central curves undershoot the data by a factor 2--3.
All in all, it is presently too early
to draw definite conclusions on the applicability of standard QCD heavy
quark calculations at RHIC.

Single inclusive electron spectra are also of great interest in relativistic
nucleus--nucleus collisions since they test ideas about energy loss of
particles in media. 
Generally, at transverse momenta of a few GeV, the energy loss of 
the heavier bottom quarks is expected to be smaller than that
of the lighter charm quarks. Accordingly, the electrons from
$b$-decays are much less suppressed.
A calculation of the suppression of single inclusive electrons
in $Au$--$Au$ collisions compared to $p p$ collisions
(quenching ratio) shows that the result depends critically on the
relative contribution of charm and bottom production, which is
affected by large perturbative uncertainties\cite{Armesto:2005mz}.
Therefore, it would be very helpful to disentangle experimentally
the $b$- and $c$-decay contributions.

\subsection{Heavy Quarks and $k_T$-Factorization\protect\cite{zotov,peters}}
We had two talks dealing with heavy quark production in the $k_T$-factorization
approach which involve a resummation of small-$x$ logarithms which become
large at high energies.
N.\ Zotov, reported on a study of HERA beauty photoproduction data
and found reasonable agreement with the data\cite{Lipatov:2006qn}.
A similar study of beauty electroproduction can be found in Ref.\ \cite{Lipatov:2006uj}.
In general, it is important to note that much progress has been made
towards a global understanding of $k_T$-factorization by working out
this picture for several processes.
This allows to study all relevant data sets with one unintegrated gluon distribution
along with a systematic study of theoretical uncertainties 
including a variation of scales. 
%
% Peters: no discussion of uncertainties
The kinematic region of very small $x$ is also a regime where the gluon density
is growing so large that gluon recombination effects are becoming relevant.
Such effects can be taken into account with the help of the
Balitsky-Kovchegov equation which 
adds a non-linear term to the linear BFKL evolution equation of the gluon.
As has been discussed by K.\ Peters at this meeting, the results 
for bottom production at the Tevatron and the LHC with and without the 
non-linear correction to the evolution are very similar such that
linear gluon evolution can be safely applied in the $k_T$-factorization 
approach\cite{peters}.

\subsection{Charmonium Production: CEM vs.\ NRQCD\protect\cite{lee}}
J.\ Lee reported on a detailed comparison of charmonium production cross sections
in the colour evaporation model (CEM) and the framework of 
non-relativistic QCD (NRQCD)\cite{Bodwin:2005hm}.
The CEM is a simple model describing the transition from a $c \bar{c}$ pair into charmonium
which has been invented about 30 years ago.
On the other hand, NRQCD is quite a general framework derived from QCD.
Therefore it is possible to 
derive relationships between the NRQCD non-perturbative matrix elements 
that follow from the model assumptions of the CEM.
Such relations do not respect the velocity-scaling rules of NRQCD and lead to 
a rather different picture of the transition of a quark anti-quark pair into
a quarkonium. Phenomenologically, these relationships are also in disagreement
with values of the matrix elements that have been extracted from the Tevatron
data for charmonium production at large transverse momenta\cite{Bodwin:2005hm}.
Finally, also a direct comparison of the CEM and NRQCD predictions with 
charmonium production data from CDF show that the CEM fits are not satisfactory
both in normalization and slope, even if multiple gluon emission effects are
included in form of $k_T$ smearing. On the other hand, NRQCD factorization
which has more free parameters than the CEM gives a satisfactory fit to the data.
For more details see\cite{lee,Bodwin:2005hm}.

\section{Experimental summary}
\label{sec:experiment}

The study of heavy flavour production in $ep$ and hadron-hadron 
collisions provides important information on 
the gluon density of the proton and a test of the understanding of many
aspects of QCD.
The measurement of heavy flavours at  
the $e^+e^-$ $B$-factories and at 
the Tevatron allows to constrain the parameters
describing CP violation and provide a wealth of information
on heavy flavour hadron spectroscopy.  In heavy ion collisions at RHIC
heavy flavour production is a vital tool for the understanding
of QCD at high densities.
At this workshop new results from the HERA-II data taking period 
were shown. 
The experiments H1 and ZEUS have upgraded many aspects of their 
detectors including those relevant for heavy flavour production.
For example, results using the newly installed ZEUS micro-vertex detector 
were presented.
The large luminosity delivered by the HERA upgrade combined
with the detector enhancements will continue to see improved
precision of heavy flavour measurements.
The HERA experiments also continue to finalise their 
remaining HERA-I measurements with many results shown as preliminary
last year presented in their final form at this workshop.
These included the measurements of charm fragmentation in photoproduction
by ZEUS, charm and beauty jets in photoproduction by H1 and the inclusive
production of charm and beauty in DIS by H1.

\subsection{$B_s$ Mixing}

\begin{figure}[ht]
    \includegraphics[width=.48\textwidth]{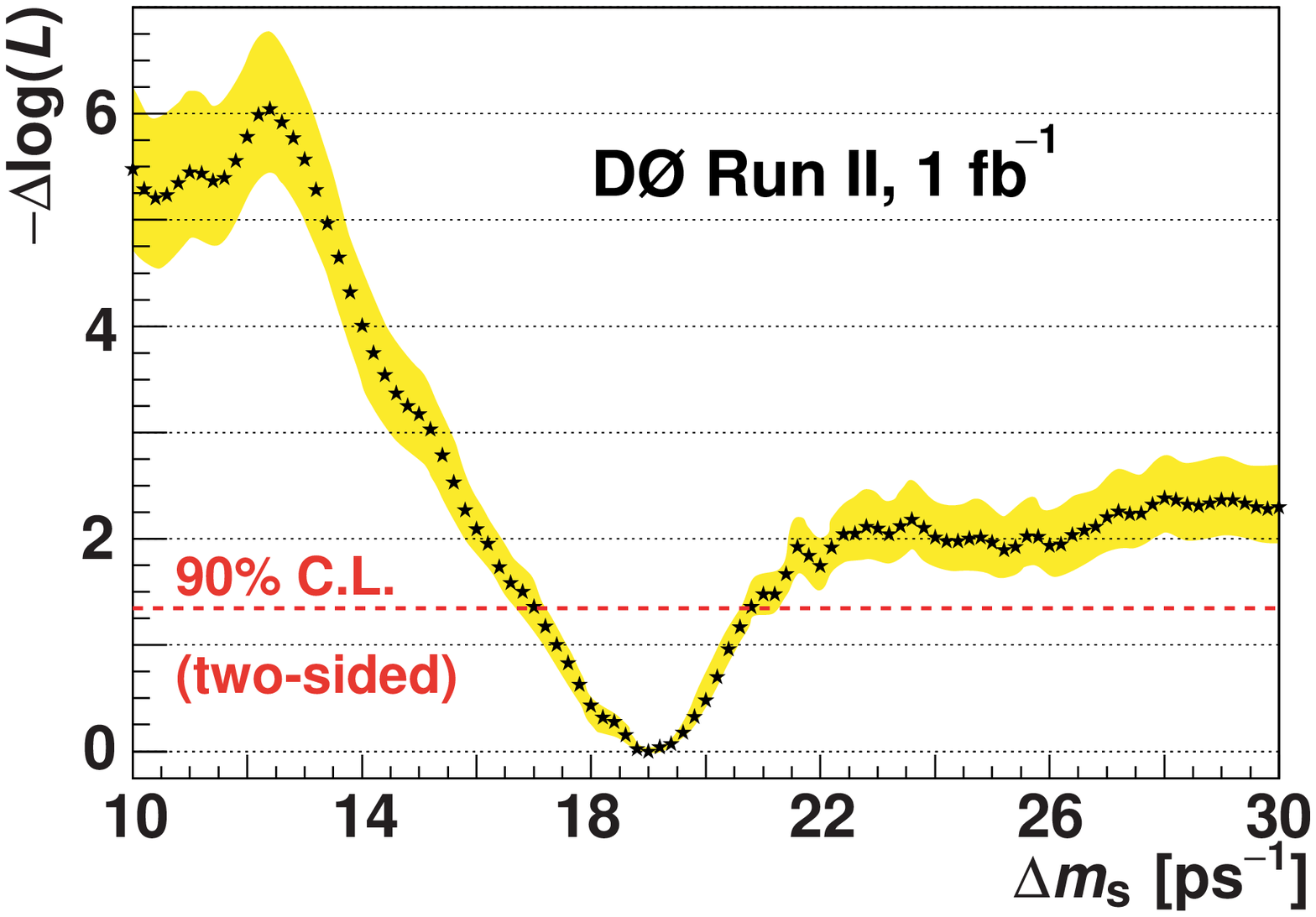}
    \includegraphics[width=.48\textwidth]{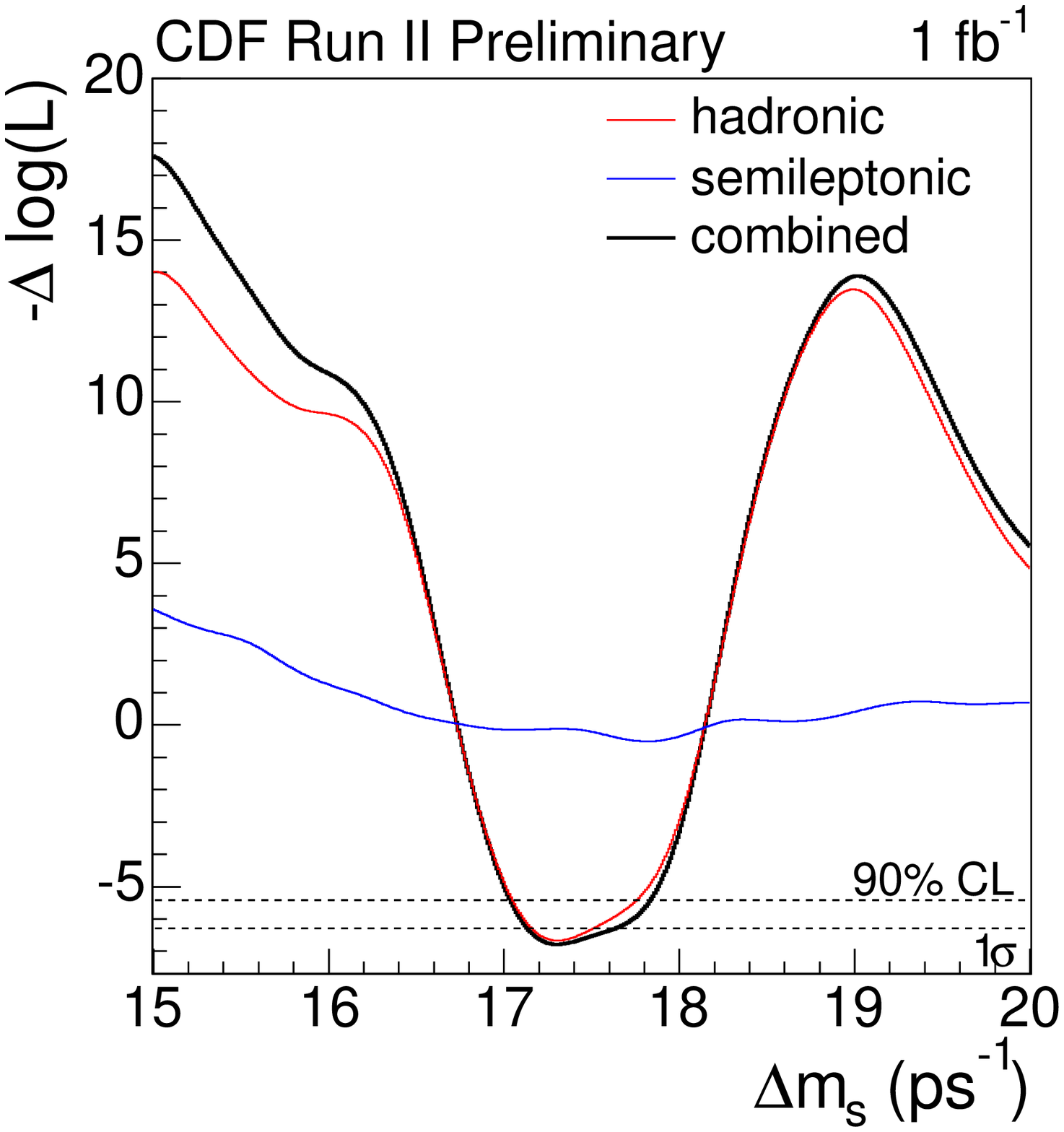}
\caption{
The likelihood for different values of the oscillation 
frequency $\Delta m_s$ shown for  D0 (left) and  CDF (right). 
\label{bsmixing}}
\end{figure}

The first measurements of the $B_s$-$\bar{B_s}$ mixing 
oscillation frequency made at the Tevatron by the
D0\cite{Abazov:2006dm} and CDF\cite{Abulencia:2006mq}
collaborations were presented by T. Kuhl.  
Mixing in the $B_s$ 
sector is only measured indirectly at the B-factories due to a low
centre of mass energy.  A determination of the oscillation 
frequency $\Delta m_s$ between the mass eigenstates 
allows the extraction of the $V_{td}$ matrix element which improves
the understanding of CP violation in the standard model.
The measurement was made possible by the $1 {\rm fb}^{-1}$
of data collected by 
each experiment.
The D0 collaboration fully reconstructed the decay of a $B_s$ meson
and tagged the opposite meson via its semi-leptonic decay.
The mixing asymmetry is measured as a function of the decay length
allowing the most likely value of $\Delta m_s$ to be extracted 
in a fit.  A value in the range $17< \Delta m_s < 21 {\rm ps}^{-1}$
at $90 \%$ confidence level was found by D0 (figure \ref{bsmixing}).  
The measurement has also
been performed by CDF which makes use of an impact parameter trigger and 
the explicit reconstruction of many hadronic and semi-leptonic 
decay channels in the
tracking detectors. A more accurate value of 
$17.33^{+0.42}_{-0.21}(stat.)\pm 0.07 (sys.t) {\rm ps}^{-1}$ is
measured (figure \ref{bsmixing}). These results on $\Delta m_s$
improve considerably the constraints on the CKM unitarity triangle. 
Further indication that the Tevatron experiments are becoming
$B$-factories themselves was provided by the talk on $B$ resonances
and $B$ hadron decays at D0 by D. Gele.

\subsection{Heavy-Quark Jets}

\begin{figure}[ht]
    \includegraphics[width=.49\textwidth]{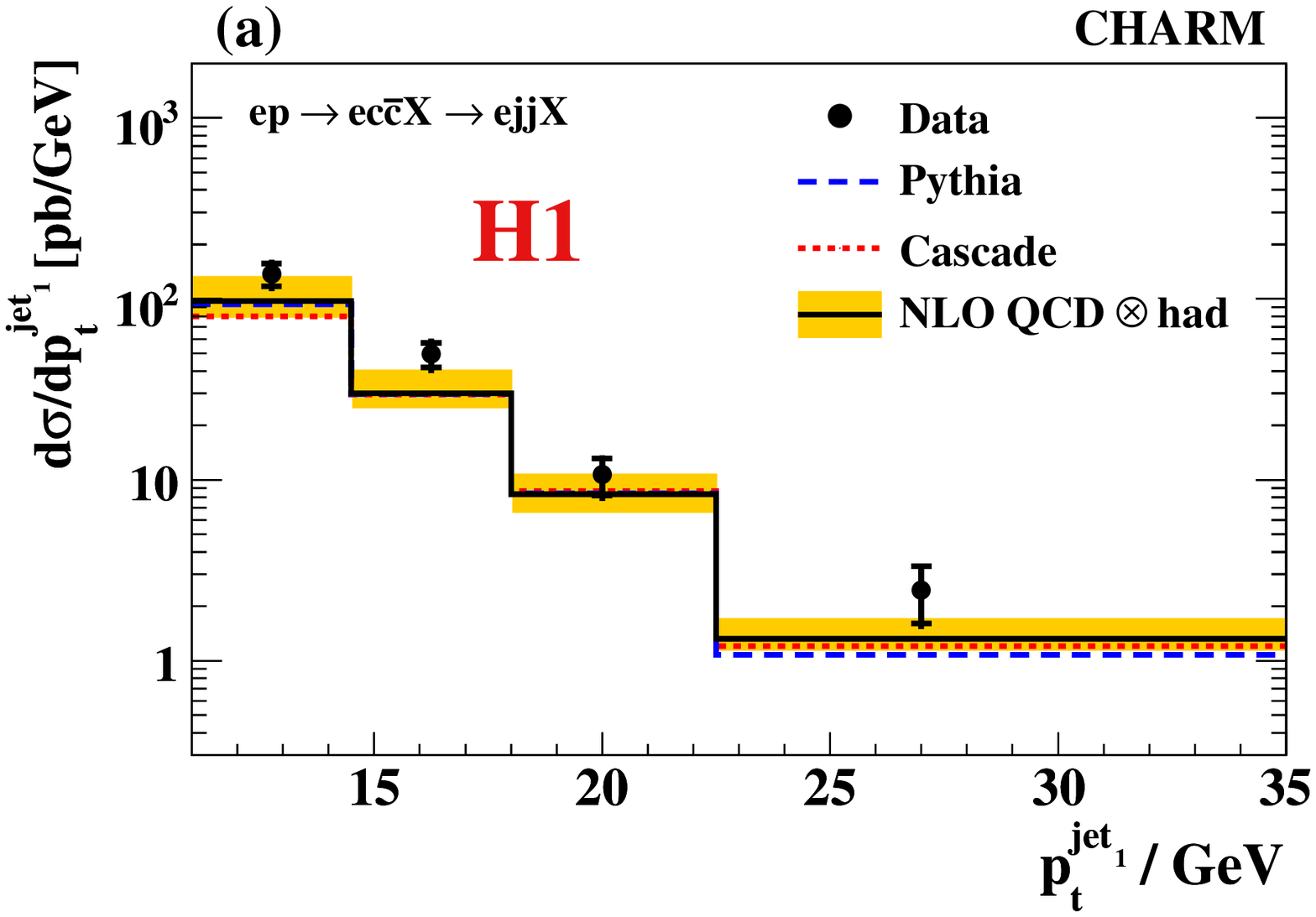}
    \includegraphics[width=.49\textwidth]{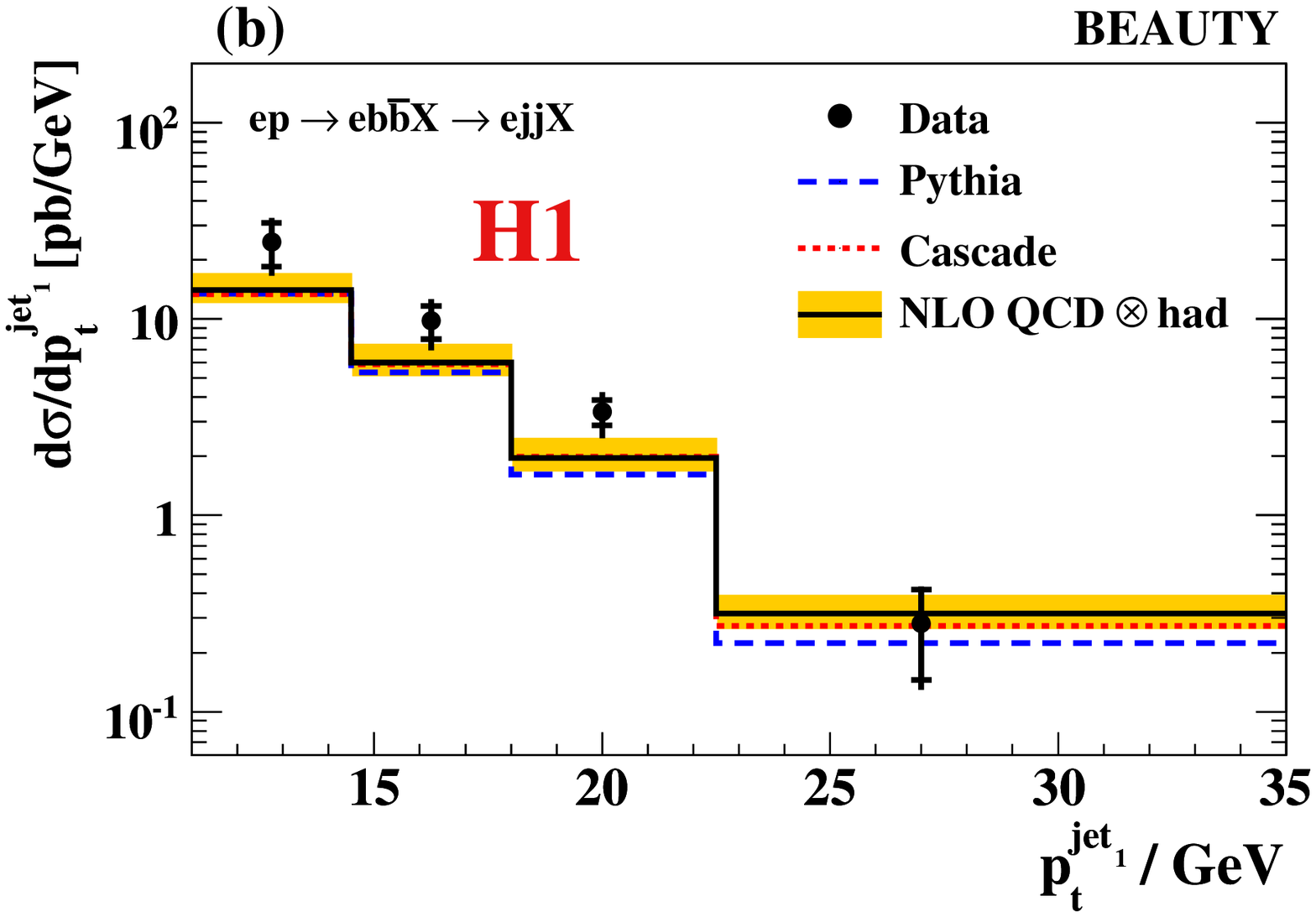}
\caption{The differential dijet cross section as a function of
$p_T$ of the highest $p_T$ jet for charm (left) and 
beauty (right) flavoured jets.
\label{h1cbjets}}
\end{figure}

Final results\cite{h1cbjets:2006vs} 
on the measurement of heavy flavour dijets in
photoproduction from H1 were shown by L. Finke. 
The cross sections are obtained by measuring the displacement
from the primary vertex of all tracks with precise spatial 
information from the H1 vertex detector. This allows the 
fraction of $c$ and $b$ components of an inclusive dijet
sample to be determined. 
The cross section is plotted as a function of $p_T$ of the
highest $p_T$ jet in figure~\ref{h1cbjets}.
Taking into account the theoretical 
uncertainties, the charm cross sections are consistent with the 
NLO QCD calculations, while the predicted cross sections for 
beauty production are somewhat lower than the measurement.

\subsection{The Inclusive Production of Heavy Flavours}

Final results\cite{Aktas:2005iw} on the inclusive production of 
charm and beauty
quarks in DIS were presented by P. Laycock. 
The measurements were made using a method based on the
displacement of tracks from the primary vertex.
The double differential reduced cross sections in $Q^2$ and $x$
are measured in the range $Q^2 > 12 \ {\rm GeV^2}$
and $0.0002 \le x \le 0.005$. 
The charm results are found to be compatible with those made using measurements
of $D^*$ cross sections where the extrapolation to the full phase 
space is larger.  The results are also found to be compatible with 
the predictions of NLO QCD although the differences in 
the predictions is as large as a factor of 2 at low $x$ and low $Q^2$. 
The HERA-II data may be able to further investigate these 
theoretical differences.

\subsection{Beauty from $D^*-\mu$ and $\mu-\mu$ Correlations}
It has long been established that the measurement of events in which 
both heavy quarks are tagged by a signature of their decay provides
large enough acceptance to measure the total
cross section with small extrapolation uncertainties. 
A. E. Nuncio Quiroz presented results from ZEUS on the 
description of $D^*-\mu$ and $\mu-\mu$  photoproduction 
cross section data by an interface of the NLO QCD program
FMNR and the Monte Carlo program PYTHIA. Comparisons of the data and theory
were, as expected, to be the same at the hadron and b-quark level.
The ZEUS $D^*-\mu$ data were found to be compatible with previous
measurements of H1 when extrapolated to the same phase space using
the new interface. The double tag correlation results continue to 
show the largest difference in the central values to NLO QCD 
(typically factor 2-3). However, the statistical significance 
of this discrepancy is still small 
due to the low tagging probabilities and more data is needed.

\subsection{Heavy Quark Production at HERA-II}

\begin{figure}[ht]
    \includegraphics[width=.49\textwidth]{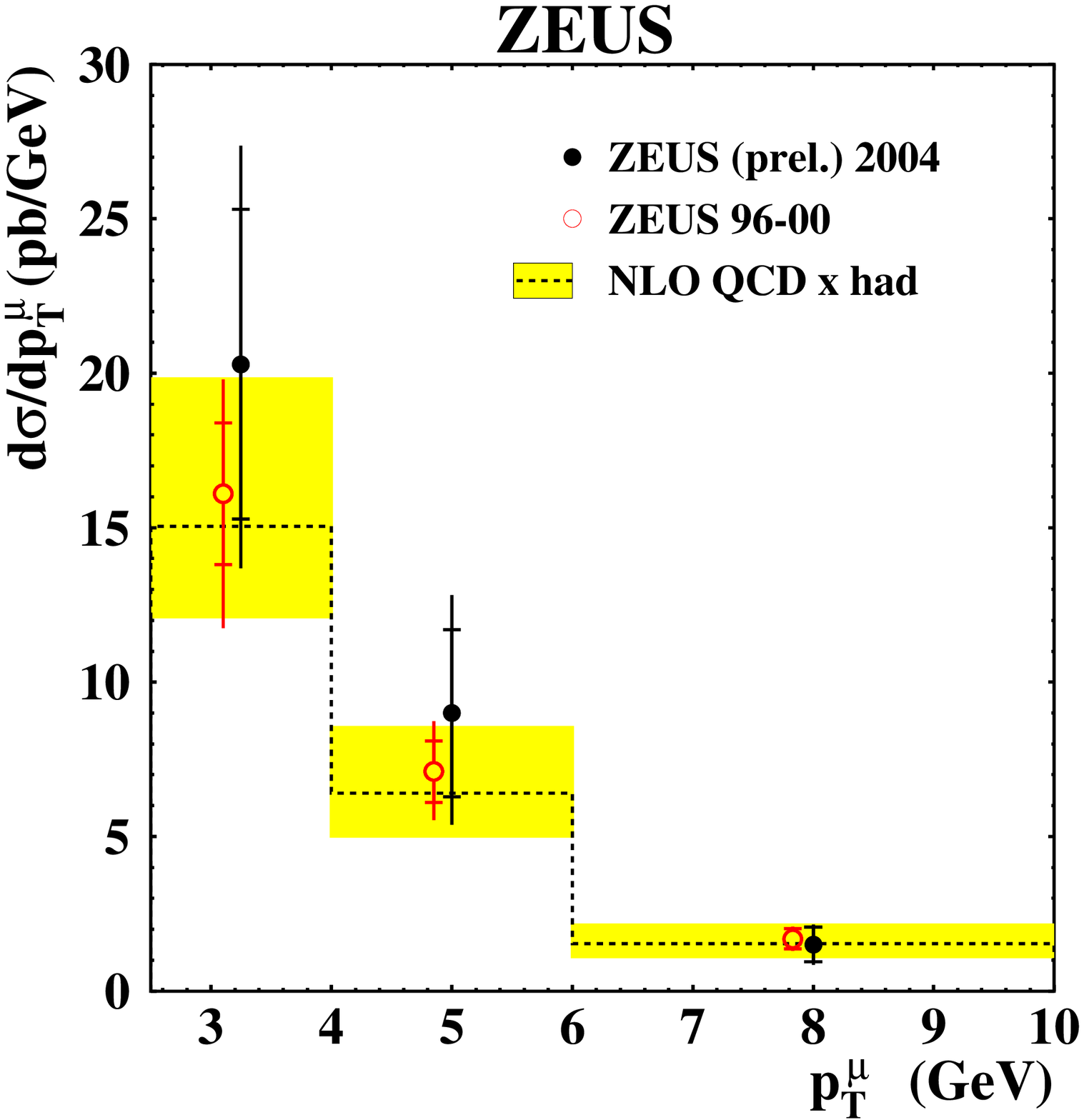}
    \includegraphics[width=.49\textwidth]{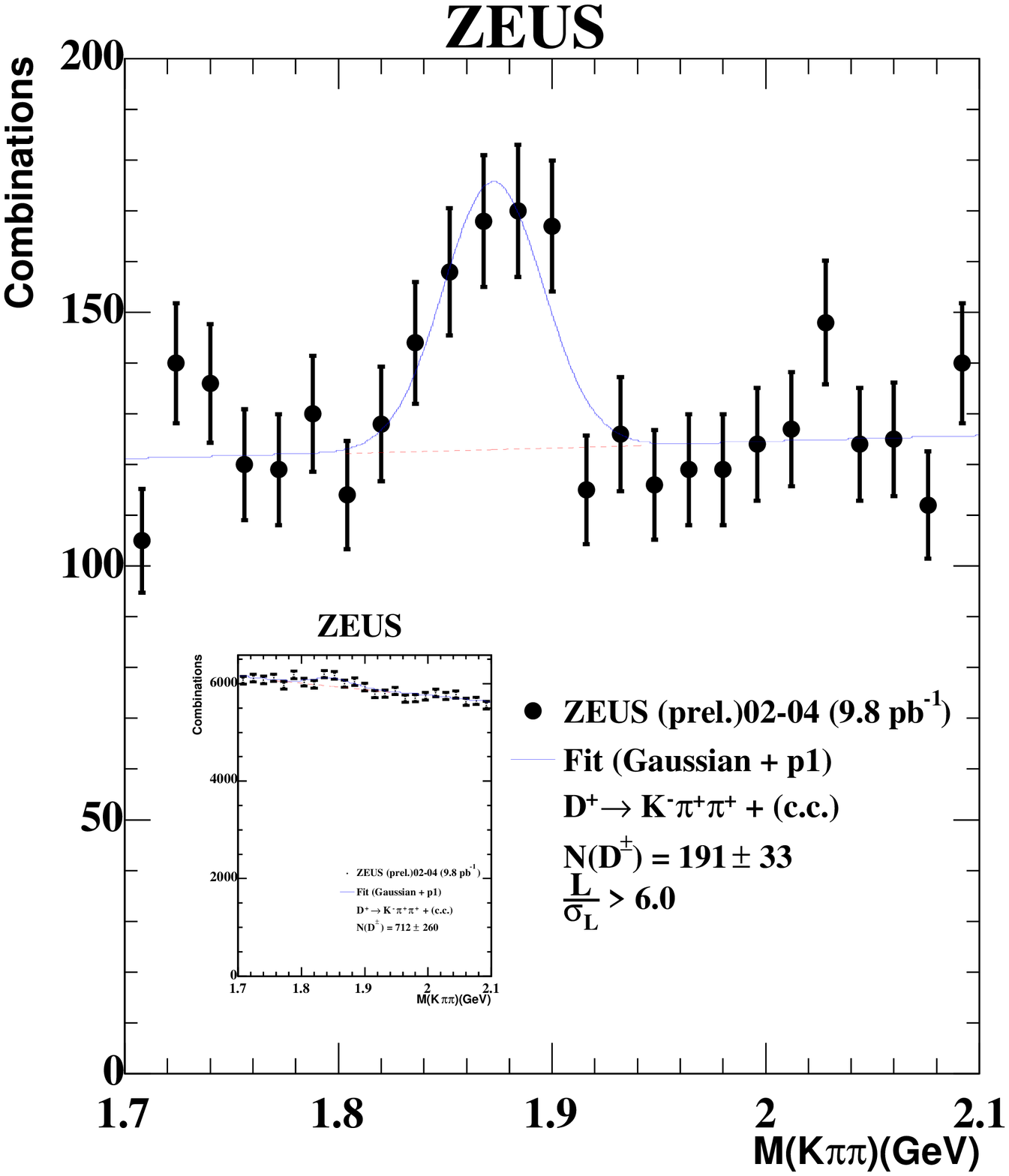}
\caption{The cross section for muon plus jets in photoproduction
as a function of the $p_T$ of the muon (left). The
invariant mass spectra for $M(K\pi\pi)$ before and
after cutting on the decay length significance of the $D^+$ 
candidate.\label{hera2}}
\end{figure}

A number of preliminary measurements from the HERA-II 
data taking period were presented at the workshop by the
ZEUS collaboration. Even though the data samples analysed
thus far represent only a small fraction of the HERA-II luminosity 
accumulated they demonstrate the technical performance of the 
detector and provide an indication of  what is feasible once the
full luminosity has been collected and analysed.  Results on beauty 
production in events with a jet and associated muon 
were presented by O. Kind. In photoproduction the
beauty cross section was obtained using a method combining
information from the relative transverse momentum of the
muon with respect to the jet ($p_T^{\rm rel}$) and the impact parameter
of the muon track as measured precisely by the micro-vertex
detector.  The cross section as a function of the
$p_T$ of the muon is shown in figure~\ref{hera2} and is seen
to be compatible with the HERA-I measurement and with 
a massive NLO QCD calculation. The measurement in DIS
was obtained using the $p_T^{\rm rel}$ method alone and
was also found to be in agreement with previous measurements
which tend to be somewhat higher than the QCD predictions.

The use of the ZEUS micro-vertex detector in charm production
was shown by F. Karstens.  The combinatorial background in the
measurement of $D^+ \rightarrow K^- \pi^+ \pi^+$ (+c.c.) can be reduced
by a factor 30 when cutting on the significance of the
decay length of the reconstructed $D^+$ meson. The invariant
mass before and after the significance cut can be seen in
figure~\ref{hera2}.

\subsection{Beauty Production at the Tevatron}

\begin{figure}[ht]
    \includegraphics[width=.49\textwidth]{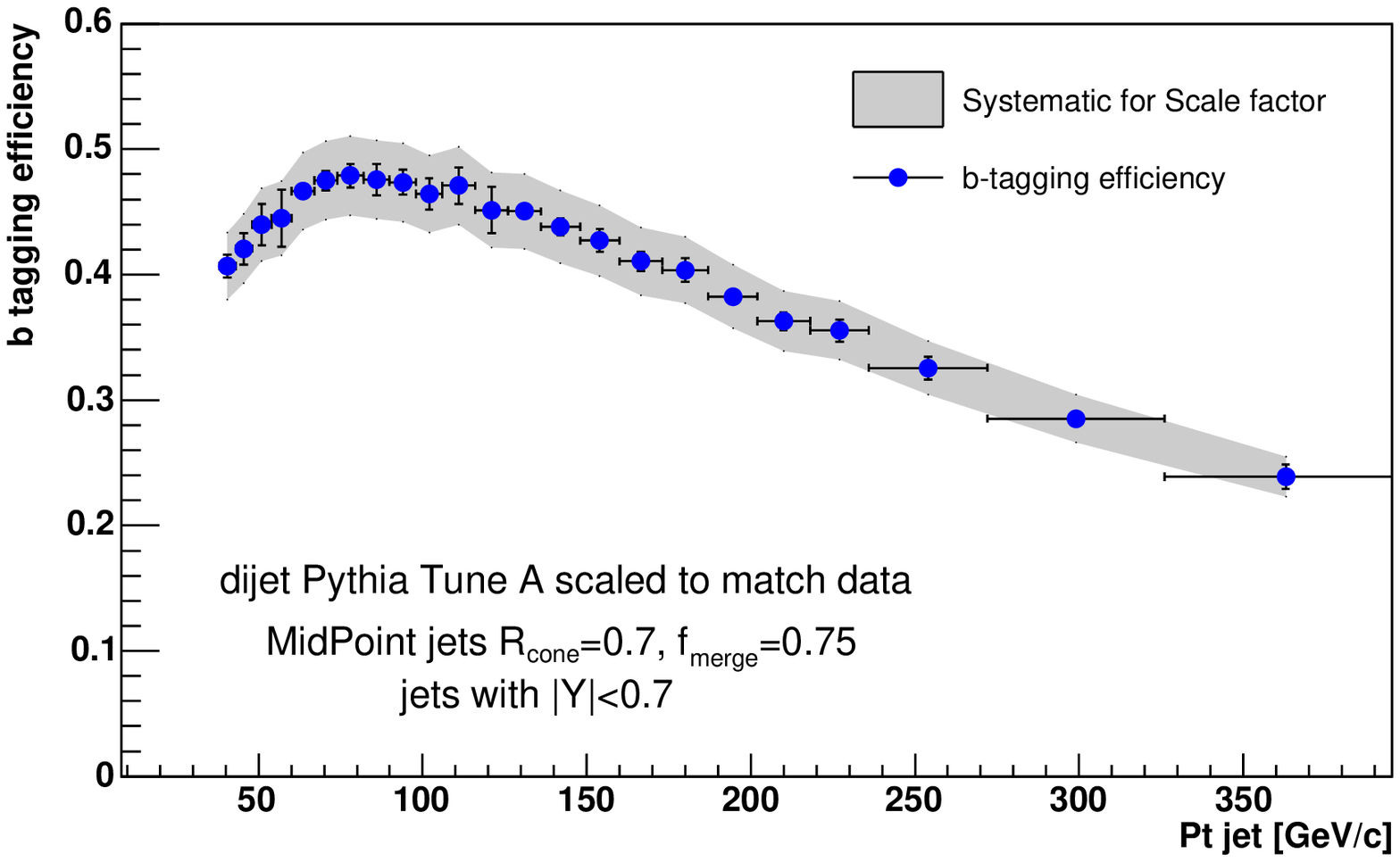}
    \includegraphics[width=.49\textwidth]{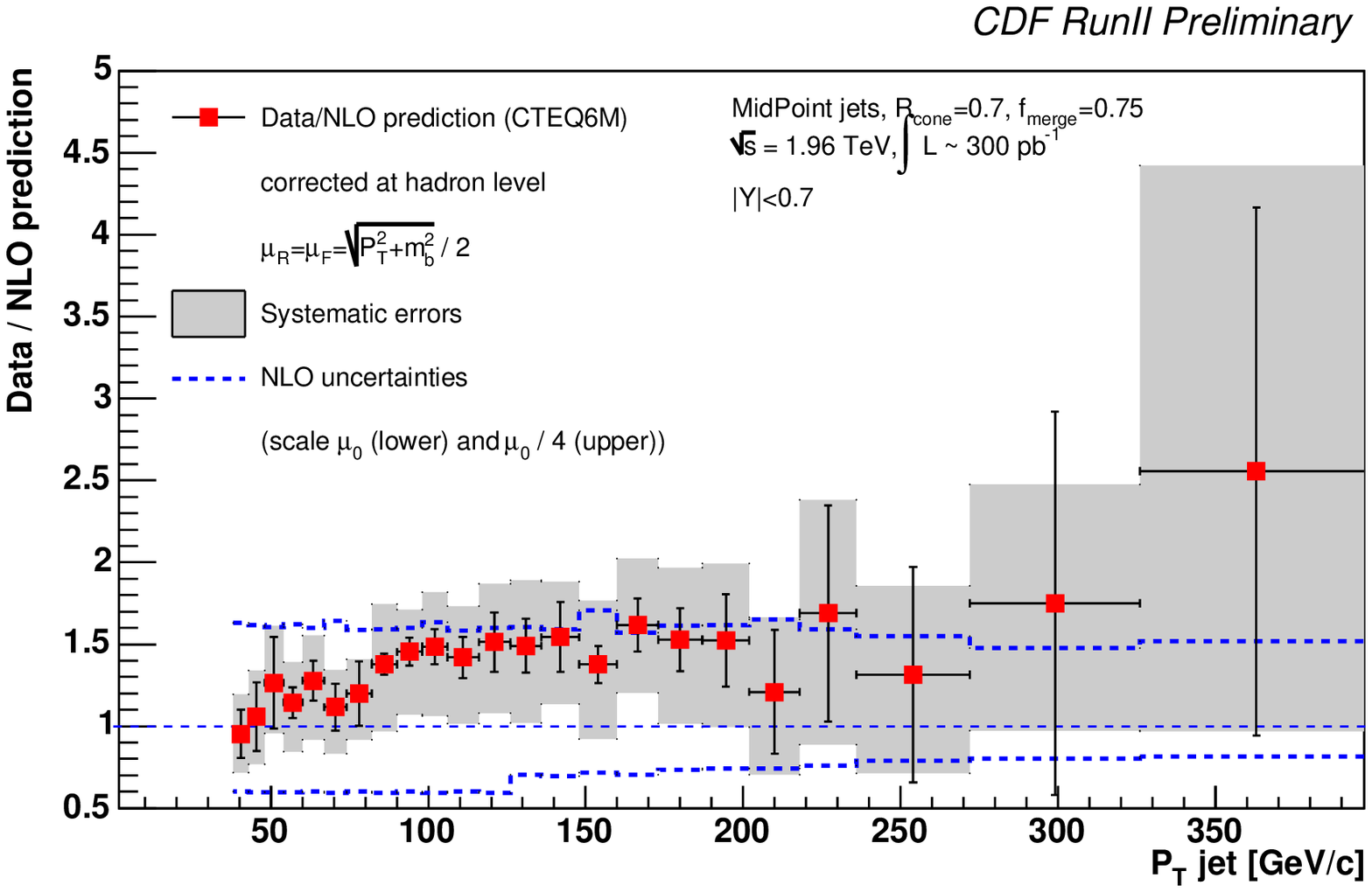}
\caption{
The efficiency of tagging $b$-jets as a function of the jet $p_T$ 
(left) and the ratio of the $b$-jet cross section
in data to a NLO QCD calculation (right).\protect\label{tevjets}}
\end{figure}

Results on the production of $b$ jets at CDF were presented
by D. Jeans.  The cross section for beauty jets was presented
as a function of $p_T$ of the jet in the range 
$38 < p_T < 400 \ {\rm GeV}$.  The
beauty component of the data is extracted by fitting the invariant
mass spectrum of particles coming from an
explicitly reconstructed secondary vertex associated with the jet.
The efficiency for tagging a $b$-jet is larger than $40\%$ for
 $p_T < 150\ {\rm GeV}$ (figure~\ref{tevjets}).  The resulting jet cross
section is found to be in agreement with a fixed order massive 
calculation in QCD.  
However, the scale uncertainties of both 
the data (due to the jet energy scale)
and the theoretical prediction
are very large.
%However, the scale uncertainty on the data
%due to the energy scale and on the theoretical prediction are
%both very large. 
These could possibly be reduced by measuring the
$b$-jet fraction.  The measurement of $Z^0$ production associated
with a tagged $b$-jet was also presented. The production
process is sensitive to the proton $b$-PDF. The CDF result
is found to be in agreement with the predictions of a  massless
NLO QCD calculation.

The status of the level of agreement between experimental data and 
theoretical predictions of $b$-production at the Tevatron was investigated
by F. Happacher.  The ratio of all published measurements to the same MC
calculation was presented.  It is observed that the ratio varies by an amount
larger than expected on statistical grounds. Therefore, a single
theory would have difficulty describing all the available data.  However, 
NLO QCD is able to describe the latest preliminary results from CDF 
and D0 (see above). Hopefully, the
final data from Run-II will help to settle the remaining
differences between the data sets.

\subsection{Charm Quark Fragmentation and Spectroscopy}

The large charm production cross section at HERA allows the study
of the fragmentation of charm quarks into the various charmed
hadrons.  Final results\cite{Chekanov:2005mm} on charm 
fragmentation in photoproduction
from ZEUS were presented by W. Dunne.  The fraction of charm
quark hadronizing into $D^0,D^+,D_s$ mesons and the $\Lambda_c$ baryon
were presented. 
The fractions are measured by explicitly
reconstructing the products in the nominal decay modes of 
%reconstructing the decay products of 
the charmed hadrons in the ZEUS central tracking detector.
The fractions are found to be compatible with %previous
measurements from H1 and ZEUS in DIS and with those from
$e^+e^-$ scattering supporting the hypothesis of a universal charm
fragmentation.  The hypothesis is further 
tested by measuring 
related quantities such as the ratio of charged to neutral
meson production, the ratio of vector meson to vector plus 
pseudoscalar meson production and the strangeness suppression factor.  
Again the range of preferred values are found to be consistent with 
those from $e^+e^-$ scattering.
L. Gladilin presented an overview of experimental
results on heavy quark fragmentation. This included
results on charm 
fragmentation fractions from $ep$ and $e^+e^-$ colliding experiments.
A summary on the latest results on the measurements of heavy flavour 
fragmentation functions which is the momentum fraction of the quark 
carried by the heavy flavour hadron were also presented.

The $B$-factories continue to provide a wealth of information on charm
spectroscopy.  The latest results form Babar, Belle and CLEO
on charm spectroscopy, charm decays and the production of new states
was presented by M. Saleem.

\subsection{Quarkonium Production}

Final results\cite{Aktas:2005xu} on the elastic production of
$J/\psi$ mesons in photoproduction from H1 were shown by Y.C. Zhu. The 
results are sensitive to different parameterisations of the gluon
density of the proton in a region where it is not constrained
by measurements of the inclusive proton structure function.
The B-factories are a source of a huge amount of 
quarkonium data.
The latest results 
on quarkonium spectroscopy, new states and charm baryons 
from Babar and Belle were presented by L. Vitale and H. Kichimi,
respectively.

\subsection{Heavy Flavour Production at RHIC and HERA-B}

The mass of charm quarks and the fact that they are 
mostly produced via gluon fusion in hadron-hadron collisions
makes them ideal probes of the dynamics of heavy ion collisions
and their description by perturbative QCD. The production of
charm at forward and backward rapidities in $p p$ and $d Au$
collisions at the PHENIX experiment was presented by X. Wang.
Charm production at forward rapidity is found to be suppressed consistent
with predictions in two different models, the colour glass condensate 
and a model based on power corrections.
However, at backward rapidity the results differ from the expectations
although more precise data is required.  Measurements from STAR
for central heavy flavour production in $Au Au$
collisions were also shown by M. Calderon. The suppression of heavy flavours
w.r.t. $p p$ collisions at high $p_T$ suggests large energy loss
similar to light quarks.  However, the results are difficult to
interpret without experimentally distinguishing $c$ and $b$ contributions
and therefore detector upgrades are being performed.
The understanding of heavy flavour suppression in an ordinary nuclear 
state is also crucial for interpreting the formation of quark-gluon plasma.
The results on $J/\psi$ production, and many other processes, 
at the high-energy fixed target experiment HERA-B were presented by R. Spighi.

\subsection{Towards the LHC}

The first of hopefully many contributions to the heavy flavour
sessions from the LHC experiments was made by C. Ciocca who gave a 
presentation on the study of top pair production at CMS.  The heavy flavour 
session was concluded by a presentation
by M. Wing illustrating the relevance of heavy flavour production 
at HERA to the LHC.  The participants of the conference were
encouraged to participate in the ongoing HERA-LHC workshop\cite{heralhc}.

%\section{Conclusions/Final comments ???}
%\label{sec:conclusions}
%
%\begin{itemize}
%\item In future more work on $b$-PDFs, intrinsic charm (Wu-Ki)
%\item Other open issues?
%\end{itemize}

\section*{Acknowledgments}
We wish to thank all the speakers for their contributions and
the organizers of the DIS06 for this perfectly organized
workshop.

%\bibliographystyle{/home/schien/Bibliography/test}
%\bibliography{/home/schien/Bibliography/heavyquarks,/home/schien/Bibliography/dis06}

\begin{thebibliography}{10}
\expandafter\ifx\csname bibnamefont\endcsname\relax
  \def\bibnamefont#1{#1}\fi
\expandafter\ifx\csname bibfnamefont\endcsname\relax
  \def\bibfnamefont#1{#1}\fi
\expandafter\ifx\csname url\endcsname\relax
  \def\url#1{\texttt{#1}}\fi
\expandafter\ifx\csname urlprefix\endcsname\relax\def\urlprefix{URL }\fi
\expandafter\ifx\csname bibinfo\endcsname\relax \def\bibinfo#1#2{#2}\fi
\expandafter\ifx\csname eprint\endcsname\relax \def\eprint#1{#1}\fi

\bibitem{thorne}
\bibinfo{note}{{R.\ Thorne, these proceedings}}.

\bibitem{tung}
\bibinfo{note}{{W.-K.\ Tung, these proceedings}}.

\bibitem{Martin:2006qz}
\bibinfo{author}{\bibfnamefont{A.~D.} \bibnamefont{Martin}},
  \bibinfo{author}{\bibfnamefont{W.~J.} \bibnamefont{Stirling}},
  \bibnamefont{and} \bibinfo{author}{\bibfnamefont{R.~S.}
  \bibnamefont{Thorne}}, \bibinfo{journal}{Phys. Lett.}
  \textbf{\bibinfo{volume}{B636}}, \bibinfo{pages}{259} (\bibinfo{year}{2006}).

\bibitem{Thorne:2006qt}
\bibinfo{author}{\bibfnamefont{R.~S.} \bibnamefont{Thorne}},
  \bibinfo{journal}{Phys. Rev.} \textbf{\bibinfo{volume}{D73}},
  \bibinfo{pages}{054019} (\bibinfo{year}{2006}).

\bibitem{Collins:1998rz}
\bibinfo{author}{\bibfnamefont{J.~C.} \bibnamefont{Collins}},
  \bibinfo{journal}{Phys. Rev.} \textbf{\bibinfo{volume}{D58}},
  \bibinfo{pages}{094002} (\bibinfo{year}{1998}).

\bibitem{Tung:2001mv}
\bibinfo{author}{\bibfnamefont{W.-K.} \bibnamefont{Tung}},
  \bibinfo{author}{\bibfnamefont{S.}~\bibnamefont{Kretzer}}, \bibnamefont{and}
  \bibinfo{author}{\bibfnamefont{C.}~\bibnamefont{Schmidt}},
  \bibinfo{journal}{J. Phys.} \textbf{\bibinfo{volume}{G28}},
  \bibinfo{pages}{983} (\bibinfo{year}{2002}).

\bibitem{Kramer:2000hn}
\bibinfo{author}{\bibfnamefont{M.}~\bibnamefont{Kr{\"a}mer}},
  \bibinfo{author}{\bibfnamefont{F.~I.} \bibnamefont{Olness}},
  \bibnamefont{and} \bibinfo{author}{\bibfnamefont{D.~E.} \bibnamefont{Soper}},
  \bibinfo{journal}{Phys. Rev.} \textbf{\bibinfo{volume}{D62}},
  \bibinfo{pages}{096007} (\bibinfo{year}{2000}).

\bibitem{Kretzer:1998ju}
\bibinfo{author}{\bibfnamefont{S.}~\bibnamefont{Kretzer}} \bibnamefont{and}
  \bibinfo{author}{\bibfnamefont{I.}~\bibnamefont{Schienbein}},
  \bibinfo{journal}{Phys. Rev.} \textbf{\bibinfo{volume}{D58}},
  \bibinfo{pages}{094035} (\bibinfo{year}{1998}).

\bibitem{Buza:1998wv}
\bibinfo{author}{\bibfnamefont{M.}~\bibnamefont{Buza} {\it et al.}},
%  \bibinfo{author}{\bibfnamefont{Y.}~\bibnamefont{Matiounine}},
%  \bibinfo{author}{\bibfnamefont{J.}~\bibnamefont{Smith}}, \bibnamefont{and}
%  \bibinfo{author}{\bibfnamefont{W.~L.} \bibnamefont{van Neerven}},
  \bibinfo{journal}{Eur. Phys. J.} \textbf{\bibinfo{volume}{C1}},
  \bibinfo{pages}{301} (\bibinfo{year}{1998}).

\bibitem{Harris:1996jx}
\bibinfo{author}{\bibfnamefont{B.~W.} \bibnamefont{Harris}},
  \bibinfo{author}{\bibfnamefont{J.}~\bibnamefont{Smith}}, \bibnamefont{and}
  \bibinfo{author}{\bibfnamefont{R.}~\bibnamefont{Vogt}},
  \bibinfo{journal}{Nucl. Phys.} \textbf{\bibinfo{volume}{B461}},
  \bibinfo{pages}{181} (\bibinfo{year}{1996}).

\bibitem{kniehl}
\bibinfo{note}{{B.\ A.\ Kniehl, these proceedings}}.

\bibitem{cacciari1}
\bibinfo{note}{{M.\ Cacciari, these proceedings}}.

\bibitem{Kniehl:2004fy}
\bibinfo{author}{\bibfnamefont{B.~A.} \bibnamefont{Kniehl} {\it et al.}},
%  \bibinfo{author}{\bibfnamefont{G.}~\bibnamefont{Kramer}},
%  \bibinfo{author}{\bibfnamefont{I.}~\bibnamefont{Schienbein}},
%  \bibnamefont{and}
%  \bibinfo{author}{\bibfnamefont{H.}~\bibnamefont{Spiesberger}},
  \bibinfo{journal}{Phys. Rev.} \textbf{\bibinfo{volume}{D71}},
  \bibinfo{pages}{014018} (\bibinfo{year}{2005}).

\bibitem{Kniehl:2005mk}
\bibinfo{author}{\bibfnamefont{B.~A.} \bibnamefont{Kniehl} {\it et al.}},
%  \bibinfo{author}{\bibfnamefont{G.}~\bibnamefont{Kramer}},
%  \bibinfo{author}{\bibfnamefont{I.}~\bibnamefont{Schienbein}},
%  \bibnamefont{and}
%  \bibinfo{author}{\bibfnamefont{H.}~\bibnamefont{Spiesberger}},
  \bibinfo{journal}{Eur. Phys. J.} \textbf{\bibinfo{volume}{C41}},
  \bibinfo{pages}{199} (\bibinfo{year}{2005}).

\bibitem{ffs}
\bibinfo{author}{\bibfnamefont{B.~A.} \bibnamefont{Kniehl}} \bibnamefont{and}
  \bibinfo{author}{\bibfnamefont{G.}~\bibnamefont{Kramer}}, \eprint{hep-ph/0607306}.

\bibitem{Kniehl:2005ej}
\bibinfo{author}{\bibfnamefont{B.~A.} \bibnamefont{Kniehl} {\it et al.}},
%  \bibinfo{author}{\bibfnamefont{G.}~\bibnamefont{Kramer}},
%  \bibinfo{author}{\bibfnamefont{I.}~\bibnamefont{Schienbein}},
%  \bibnamefont{and}
%  \bibinfo{author}{\bibfnamefont{H.}~\bibnamefont{Spiesberger}},
  \bibinfo{journal}{Phys. Rev. Lett.} \textbf{\bibinfo{volume}{96}},
  \bibinfo{pages}{012001} (\bibinfo{year}{2006}).

\bibitem{Collins:1986mp}
\bibinfo{author}{\bibfnamefont{J.~C.} \bibnamefont{Collins}} \bibnamefont{and}
  \bibinfo{author}{\bibfnamefont{W.-K.} \bibnamefont{Tung}},
  \bibinfo{journal}{Nucl. Phys.} \textbf{\bibinfo{volume}{B278}},
  \bibinfo{pages}{934} (\bibinfo{year}{1986}).

\bibitem{Cacciari:2005ry}
\bibinfo{author}{\bibfnamefont{M.}~\bibnamefont{Cacciari}},
  \bibinfo{author}{\bibfnamefont{P.}~\bibnamefont{Nason}}, \bibnamefont{and}
  \bibinfo{author}{\bibfnamefont{C.}~\bibnamefont{Oleari}},
  \bibinfo{journal}{JHEP} \textbf{\bibinfo{volume}{10}}, \bibinfo{pages}{034}
  (\bibinfo{year}{2005}).

\bibitem{oleari}
\bibinfo{note}{{C.\ Oleari, these proceedings}}.

\bibitem{Cacciari:2005uk}
\bibinfo{author}{\bibfnamefont{M.}~\bibnamefont{Cacciari}},
  \bibinfo{author}{\bibfnamefont{P.}~\bibnamefont{Nason}}, \bibnamefont{and}
  \bibinfo{author}{\bibfnamefont{C.}~\bibnamefont{Oleari}},
  \bibinfo{journal}{JHEP} \textbf{\bibinfo{volume}{04}}, \bibinfo{pages}{006}
  (\bibinfo{year}{2006}).

\bibitem{cacciari2}
\bibinfo{note}{{M.\ Cacciari, these proceedings}}.

\bibitem{Cacciari:2005rk}
\bibinfo{author}{\bibfnamefont{M.}~\bibnamefont{Cacciari}},
  \bibinfo{author}{\bibfnamefont{P.}~\bibnamefont{Nason}}, \bibnamefont{and}
  \bibinfo{author}{\bibfnamefont{R.}~\bibnamefont{Vogt}},
  \bibinfo{journal}{Phys. Rev. Lett.} \textbf{\bibinfo{volume}{95}},
  \bibinfo{pages}{122001} (\bibinfo{year}{2005}).

\bibitem{Cacciari:1998it}
\bibinfo{author}{\bibfnamefont{M.}~\bibnamefont{Cacciari}},
  \bibinfo{author}{\bibfnamefont{M.}~\bibnamefont{Greco}}, \bibnamefont{and}
  \bibinfo{author}{\bibfnamefont{P.}~\bibnamefont{Nason}},
  \bibinfo{journal}{JHEP} \textbf{\bibinfo{volume}{05}}, \bibinfo{pages}{007}
  (\bibinfo{year}{1998}).

\bibitem{Armesto:2005mz}
\bibinfo{author}{\bibfnamefont{N.}~\bibnamefont{Armesto} {\it et al.}},
%  \bibinfo{author}{\bibfnamefont{M.}~\bibnamefont{Cacciari}},
%  \bibinfo{author}{\bibfnamefont{A.}~\bibnamefont{Dainese}},
%  \bibinfo{author}{\bibfnamefont{C.~A.} \bibnamefont{Salgado}},
%  \bibnamefont{and} \bibinfo{author}{\bibfnamefont{U.~A.}
%  \bibnamefont{Wiedemann}}, 
\bibinfo{journal}{Phys. Lett.}
  \textbf{\bibinfo{volume}{B637}}, \bibinfo{pages}{362} (\bibinfo{year}{2006}).

\bibitem{zotov}
\bibinfo{note}{{N.\ Zotov, these proceedings}}.

\bibitem{peters}
\bibinfo{note}{{K.\ Peters, these proceedings}}.

\bibitem{Lipatov:2006qn}
\bibinfo{author}{\bibfnamefont{A.~V.} \bibnamefont{Lipatov}} \bibnamefont{and}
  \bibinfo{author}{\bibfnamefont{N.~P.} \bibnamefont{Zotov}},
  \bibinfo{journal}{Phys. Rev.} \textbf{\bibinfo{volume}{D73}},
  \bibinfo{pages}{114018} (\bibinfo{year}{2006}).

\bibitem{Lipatov:2006uj}
\bibinfo{author}{\bibfnamefont{A.~V.} \bibnamefont{Lipatov}} \bibnamefont{and}
  \bibinfo{author}{\bibfnamefont{N.~P.} \bibnamefont{Zotov}},
%  \emph{\bibinfo{title}{{Deep inelastic beauty production at HERA in the
%  $k_T$-factorization approach}}} (\bibinfo{year}{2006}),
  \eprint{hep-ph/0603017}.

\bibitem{lee}
\bibinfo{note}{{J.\ Lee, these proceedings}}.

\bibitem{Bodwin:2005hm}
\bibinfo{author}{\bibfnamefont{G.~T.} \bibnamefont{Bodwin}},
  \bibinfo{author}{\bibfnamefont{E.}~\bibnamefont{Braaten}}, \bibnamefont{and}
  \bibinfo{author}{\bibfnamefont{J.}~\bibnamefont{Lee}},
  \bibinfo{journal}{Phys. Rev.} \textbf{\bibinfo{volume}{D72}},
  \bibinfo{pages}{014004} (\bibinfo{year}{2005}).

\bibitem{Abazov:2006dm}
  V.~M.~Abazov {\it et al.}  [D0 Collaboration],
  %``First direct two-sided bound on the B/s0 oscillation frequency,''
  Phys.\ Rev.\ Lett.\  {\bf 97} (2006) 021802.
%  [hep-ex/0603029].
  %%CITATION = HEP-EX 0603029;%%

\bibitem{Abulencia:2006mq}
  A.~Abulencia  [CDF - Run II Collaboration],
  %``Measurement of the B/s0 anti-B/s0 oscillation frequency,''
  [hep-ex/0606027].
  %%CITATION = HEP-EX 0606027;%%

\bibitem{h1cbjets:2006vs}
 A.~Aktas {\it et al.} 
    [H1 Collaboration],
  %``Measurement of charm and beauty dijet cross sections in photoproduction at
  %HERA using the H1 vertex detector,''
  [hep-ex/0605016].
  %%CITATION = HEP-EX 0605016;%%

\bibitem{Aktas:2005iw}
  A.~Aktas {\it et al.}  [H1 Collaboration],
  % ``Measurement of F2(c anti-c) and F2(b anti-b) at low Q**2 and x using  the
  %H1 vertex detector at HERA,''
  % Eur.\ Phys.\ J.\ C {\bf 45} (2006) 23
  [hep-ex/0507081].
  %%CITATION = HEP-EX 0507081;%%


\bibitem{Chekanov:2005mm}
  S.~Chekanov {\it et al.}  [ZEUS Collaboration],
  %``Measurement of charm fragmentation ratios and fractions in  photoproduction
  %at HERA,''
  Eur.\ Phys.\ J.\ C {\bf 44} (2005) 351.
%  [hep-ex/0508019].
  %%CITATION = HEP-EX 0508019;%%
  %%Cited 6 times in SPIRES-HEP

\bibitem{Aktas:2005xu}
  A.~Aktas {\it et al.}  [H1 Collaboration],
  %``Elastic J/psi production at HERA,''
  [hep-ex/0510016].
  %%CITATION = HEP-EX 0510016;%%

\bibitem{heralhc}
``The HERA-LHC Workshop'',
{\verb+http://www.desy.de/~heralhc/+}

\end{thebibliography}

\end{document}